\documentclass[journal,draftclsnofoot,onecolumn]{IEEEtran}

\usepackage{amssymb}

\usepackage{amsmath}

\usepackage{cite} 

\usepackage[square, comma, sort&compress, numbers]{natbib}

\begin{document}

\newtheorem{theorem}{Theorem}
\newtheorem{proposition}{Proposition}
\newtheorem{lemma}{Lemma}
\newtheorem{definition}{Definition}
\newtheorem{corollary}{Corollary}
\newtheorem{remark}{Remark}
\newtheorem{example}{Example}

\title{Some  results about permutation properties of a kind of binomials over finite fields
 \thanks{ This work is supported in part by "Funding for scientific research start-up" of Nanjing Tech University.} }

\author{Xiaogang~Liu  
\thanks{
X. Liu is with
College of Computer Science and Technology,
Nanjing Tech University, 
Nanjing City,
Jiangsu Province,
PR China
211800
     e-mail:liuxg0201@163.com. }
}

\maketitle

\begin{abstract}
Permutation polynomials have many applications in finite fields theory, coding theory, cryptography, combinatorial design, communication theory, and so on. Permutation binomials of the form $x^{r}(x^{q-1}+a)$ over $\mathbb{F}_{q^2}$ have been studied before,   K. Li, L. Qu and X. Chen proved that they are permutation polynomials if and only if $r=1$ and $a^{q+1}\not=1$. In this paper, we consider the same binomial, but over finite fields $\mathbb{F}_{q^3}$ and $\mathbb{F}_{q^e}$. Two different kinds of methods are employed, and some partial results are obtained for them.
\end{abstract}

%\begin{IEEEkeywords}
%exponential sums, finite fields, binomials, permutation polynomials
   
%\end{IEEEkeywords}

\section{Introduction}\label{secI}
 
Let $\mathbb{F}_q$ be a finite field with $q$ elements where $q$ is a power of a prime. A polynomial $f: x\rightarrow f(x)$ from $\mathbb{F}_q$ to $\mathbb{F}_q$ is called a permutation polynomial (PP) if it is both injetive and surjective. They are active research objectives due to their applications in coding, cryptography, combinatorial design, etc. So, from both theoretical and application aspects, the study of permutation polynomials is  interesting. The study of permutation polynomials have a long history, in fact, Hermite and Dickson already investigated them \cite{A00091,A00014}. A wealth of results are obtained, and some recent results are \cite{A0001,A0002,A0003,A0005,A0006,A0007,F001,A00013,A00015,A00023,A00019,A00025,A00027,Z001}.

Permutaion polynomials with few terms have simple algebraic forms and some extraordinary properties, they attract people's interest especially. For the construction and investigation of permutation binomials and trinomials of recent work, we refer the reader to \cite{A0007,A0009,A00011,A000152,A00012,A00018,A00020}, and the references therein. Carlitz studied permutation binomial in 1962 \cite{A0003c}; for fixed $d$ and large $q$,  Carlitz and Wells showed that $f=x(x^{{q-1}\over d} + a)$ can be a permutation polynomial \cite{A0003cw}. When $q\geq (m^2-4m+6)^2$, Niederreiter and Robinson proved that $f=x^m+ax\in \mathbb{F}_q[x]$ is not a permutation binomial over $\mathbb{F}_q$ \cite{A00033}. Necessary and sufficient conditions are given by Wang for $x^n(x^{{q-1}\over d}+1)$ to be a permutation binomial over $\mathbb{F}_q$ \cite{A00034}. Using Hermite's Criterion, Hou and Lappano gave several permutation binomials of the form $ax+x^{sq-s+1}$ \cite{A000152,A00012}. Also, if $f(x)$ is a complete permutation monomial, then $f(x)+x$ is a permutation binomial over $\mathbb{F}_q$.

In \cite{A00018}, the authors showed that $f(x)=x^r(x^{q-1}+a)$ is a permutation polynomial over $\mathbb{F}_{q^2}$ if and only $r=1$ and $a^{q+1}\not=1$. The purpose of this paper is to study the permutation property of $f(x)$ over $\mathbb{F}_{q^3}$ and $\mathbb{F}_{q^e}$. For the field $\mathbb{F}_{q^3}$,   Hermite's Criterion is used for this in Section II,  and it is found that $f(x)$ is almost always not a permutation polynomial, and we guess that $f(x)$ can be  a permutation polynomial  only when $r=1$ for odd characteristic. For the field $\mathbb{F}_{q^e}$, we showed that for $e$ large enough, $f(x)$ is not a permutation polynomial, and Hasse-Weil bound is used to verify this in Section III.

  The following lemmas are included in different formats in the literature, they will be useful for our discussion.

\begin{lemma}\label{lemma1}(\cite{A00026,A00034,A00045})
Let $d,r>0$ with $d\mid q-1$, and let $h(x)\in \mathbb{F}_q[x]$. Then
$f(x)=x^rh(x^{(q-1)/d})$ permutes $\mathbb{F}_q$ if and only if the following two conditions hold:

\begin{enumerate}
\renewcommand{\labelenumi}{$($\mbox{\roman{enumi}}$)$}
\item
$ \textup{gcd}(r,(q-1)/d)=1;$
\item
$x^rh(x)^{(q-1)/d}$ permutes $\mu_{d}, $ where $\mu_{d}$ denotes the $d$-th root of unity in $\mathbb{F}_q$.
 
\end{enumerate}

\end{lemma}

\begin{lemma}\label{lemma2}(Hermite Criterion \cite{A00014,A000242})
Let $\mathbb{F}_q$ be of characteristic $p$. Then $f(x)\in \mathbb{F}_q[x]$ is a permutation polynmoial of $\mathbb{F}_q$ if and only if the following two conditions hold:
\begin{enumerate}
\renewcommand{\labelenumi}{$($\mbox{\roman{enumi}}$)$}
\item
$f$ has exactly one root in $\mathbb{F}_q$;
\item
for each integer $t$ with $1\leq t\leq q-2$ and $t\not= 0 \ \ \textup{mod}\ p$, the reduction of $f(x)^t\ \textup{mod}\ x^q-x$ has degree $\leq q-2$.
 
\end{enumerate}

\end{lemma}
Note that, if $f(x)^t\ \textup{mod}\ x^q-x$ has degree $\leq q-2$, then $\sum\limits_{x\in \mathbb{F}_q} f(x)^t=0$.
 
\begin{lemma}\label{lemma3}(Lucas formula)
Let $n,i$ be positive integers and $p$ be a prime. Assume $n=a_mp^m+\cdots+a_1p+a_0$ and $i= b_mp^m+\cdots+b_1p+b_0$. Then 
\[
\dbinom{n}{i}\ \equiv \ \dbinom{a_m}{b_m}\dbinom{a_{m-1}}{b_{m-1}}\cdots\dbinom{a_0}{b_0} \ (\textup{mod}\ p).
\]
\end{lemma}
Here the binomial coefficients $ \dbinom{a_i}{b_i}$ is zero when $a_i<b_i$ for some $i$.

\section{The case of finite field  $\mathbb{F}_{q^3}$}

In this section, for $a\in \mathbb{F}_{q^3}^*$, we study polynomial $f(x)=x^r(x^{q-1}+a)$, and find that it is almost always not a permutation polynomial, except for the case that $r=1$.

Since only some partial results are obtained for the permutation behavior of $f(x)$. Before stating and proving the following Theorem \ref{t01},   a general analysis is given as how to use Hermite's Criterion to verify that $f(x)$ is not a permutation polynomial, expecting that a complete answer can be given. In fact, our aim is to find an integer  $0\leq N\leq q^3-1$ such that
\[
\sum\limits_{x\in \mathbb{F}_{q^3}}f(x)^N\not= 0.
\]

 For every integer $0\leq N\leq q^3-1$, there is a unique expression of the form $N=\alpha +\beta q+\gamma q^2$ with $0\leq \alpha,\beta,\gamma\leq q-1$. Note that, according to Lemma \ref{lemma2}, it is not necessary to consider the cases of $\alpha=\beta=\gamma=0, \alpha=\beta=\gamma=q-1$, thus $1\leq \alpha+\beta+\gamma\leq 3q-2$. Then
\begin{equation}\label{eq001}
\begin{array}{lll}
\sum\limits_{{x\in \mathbb{F}_{q^3}}}f(x)^{\alpha+\beta q+\gamma q^2}&=&
          \sum\limits_{{x\in \mathbb{F}_{q^3}}}x^{r({\alpha+\beta q+\gamma q^2})}(x^{q-1}+a)^{{\alpha+\beta q+\gamma q^2}}\\
                  &=&
\sum\limits_{{x\in \mathbb{F}_{q^3}^*}}x^{r({\alpha+\beta q+\gamma q^2})}(x^{q-1}+a)^{{\alpha }}(x^{q^2-q}+a^q)^{{\beta}}(x^{1-q^2}+a^{q^2})^{{\gamma}}\\
                                  &=&
\sum\limits_{{x\in \mathbb{F}_{q^3}^*}}x^{r({\alpha+\beta q+\gamma q^2})}\sum\limits_{i=0}^{\alpha} \dbinom{\alpha}{i} x^{(q-1)i}a^{\alpha-i}\sum\limits_{j=0}^{\beta}\dbinom{\beta}{j}  x^{(q^2-q)j}{(a^q)^{\beta-j}}\sum\limits_{k=0}^{\gamma} \dbinom{\gamma}{k} x^{(1-q^2)k}{(a^{q^2})}^{\gamma-k}\\ 
                                  &=&
a^{\alpha+\beta q+\gamma q^2}\sum\limits_{0\leq i\leq \alpha, 0\leq j\leq \beta, 0\leq k\leq \gamma}   \dbinom{\alpha}{i}\dbinom{\beta}{j}\dbinom{\gamma}{k}a^{-i-qj-q^2k} 
\sum\limits_{{x\in \mathbb{F}_{q^3}^*}}x^{r({\alpha+\beta q+\gamma q^2})+(q-1)i+(q^2-q)j+(1-q^2)k}  %x^{(q-1)i}a^{\alpha-i}\sum\limits_{j=0}^{\beta}  x^{(q^2-q)j}{(a^q)^{\beta-j}}\sum\limits_{k=0}^{\gamma}  x^{(1-%q^2)k}{(a^{q^2})}^{\gamma-k}  
.
 \end{array}
\end{equation}
The inner sum is zero, unless 
\begin{equation*}\label{eq02}
{r({\alpha+\beta q+\gamma q^2})+(q-1)i+(q^2-q)j+(1-q^2)k} \equiv 0 \ \ (\textup{mod} \ \ q^3-1). 
\end{equation*}
But 
%\[
%{%r({\alpha+\beta q+\gamma q^2})+(q-1)i+(q^2-q)j+(1-q^2)k}\\
%={r(({\alpha+\beta+\gamma)+\beta (q-1)+\gamma (q^2-1)})+(q-1)i+(q^2-q)j+(1-q^2)k}\\
%=r({\alpha+\beta+\gamma)}+{{(r(\beta +\gamma (q+1))}+i+qj-(q+1)k)(q-1)}
%\]

\begin{equation*}\label{eq01}
\begin{array}{lll}
&& {r({\alpha+\beta q+\gamma q^2})+(q-1)i+(q^2-q)j+(1-q^2)k}\\
&=&  {r(({\alpha+\beta+\gamma)+\beta (q-1)+\gamma (q^2-1)})+(q-1)i+(q^2-q)j+(1-q^2)k}\\   
&=& r({\alpha+\beta+\gamma)}+{{(r(\beta +\gamma (q+1))}+i+qj-(q+1)k)(q-1)},     
 \end{array}
\end{equation*}
which is congruent to $0$ modulo $q^3-1$ only when $r({\alpha+\beta+\gamma })$ is divisible by $q-1$, as $q-1$ is both a divisor of $q^3-1$ and $(r(\beta +\gamma (q+1))+i+qj-(q+1)k)(q-1)$.
Since $q-1=(q^3-1)/{(q^2+q+1)}$, that is  $d=q^2+q+1$ in Lemma \ref{lemma1}, we only have to consider the case that $\textup{gcd}(r,q-1)=1$. %\ \ \ \quad $r_0q+1=r_0(q-1)+r_0+1$ and 
There are two possibilities: $\alpha+\beta+\gamma=q-1$ or $\alpha+\beta+\gamma=2(q-1)$. 
%And for every integer $0\leq N\leq q^3-1$, there is a unique expression of the form $N=\alpha +\beta q+\gamma q^2$ with $0\leq \alpha,\beta,\gamma\leq q-1$. Note that, according to Lemma \ref{lemma2} we do not consider the cases of $\alpha=\beta=\gamma=0, q-1$, and $1\leq \alpha+\beta+\gamma\leq 3q-2$.

Let us assume that $\alpha+\beta+\gamma=2(q-1)$, that is $\gamma=2(q-1)-\alpha-\beta$. Then equation (\ref{eq001}) becomes
\begin{equation}\label{eq002}
\begin{array}{lll}
&&\sum\limits_{{x\in \mathbb{F}_{q^3}}}f(x)^{\alpha+\beta q+\gamma q^2}\\

&=&
         a^{\alpha+\beta q+(2(q-1)-\alpha-\beta) q^2}\sum\limits_{0\leq i\leq \alpha, 0\leq j\leq \beta, 0\leq k\leq 2(q-1)-\alpha-\beta}   \dbinom{\alpha}{i}\dbinom{\beta}{j}\dbinom{2(q-1)-\alpha-\beta}{k}a^{-i-qj-q^2k} \\
 &&
\sum\limits_{{x\in \mathbb{F}_{q^3}^*}}x^{r({\alpha+\beta q+(2(q-1)-\alpha-\beta) q^2})+(q-1)i+(q^2-q)j+(1-q^2)k} \\
&=&
         a^{(1-q)((2+\alpha)(1+q)+\beta q)}\sum\limits_{0\leq i\leq \alpha, 0\leq j\leq \beta, 0\leq k\leq 2(q-1)-\alpha-\beta}   \dbinom{\alpha}{i}\dbinom{\beta}{j}\dbinom{2(q-1)-\alpha-\beta}{k}a^{-i-qj-q^2k} \\
 &&
\sum\limits_{{x\in \mathbb{F}_{q^3}^*}}x^{[r(2q^2-(q+1)(\alpha+\beta)+\beta)+i+qj-(q+1)k](q-1)} \\
&=&
         -a^{(1-q)((2+\alpha)(1+q)+\beta q)}\sum\limits_{{[r(2q^2-(q+1)(\alpha+\beta)+\beta)+i+qj-(q+1)k](q-1)}\equiv 0\ (\textup{mod}\ q^3-1)}   \dbinom{\alpha}{i}\dbinom{\beta}{j}\dbinom{2(q-1)-\alpha-\beta}{k}a^{-i-qj-q^2k} \\
 &=&
         -a^{(1-q)((2+\alpha)(1+q)+\beta q)}\sum\limits_{{r(2q^2-(q+1)(\alpha+\beta)+\beta)+i+qj-(q+1)k}\equiv 0\ (\textup{mod}\ q^2+q+1)}   \dbinom{\alpha}{i}\dbinom{\beta}{j}\dbinom{2(q-1)-\alpha-\beta}{k}a^{-i-qj-q^2k}. \\

 \end{array}
\end{equation}
Now, let us consider the following equation 
\[
{{r(2q^2-(q+1)(\alpha+\beta)+\beta)+i+qj-(q+1)k}\equiv 0\ (\textup{mod}\ q^2+q+1)}.
\]
Since $\textup{gcd}(q,q^2+q+1)=1$, we can multiply $q$ on both sides of the above equation, and get the following equivalent equation
\[
{{r(2q^3-(q^2+q)(\alpha+\beta)+q\beta)+qi+q^2j-(q^2+q)k}\equiv 0\ (\textup{mod}\ q^2+q+1)}.
\]
Since $q^3 \equiv 1 \textup{mod} \ q^2+q+1$ and $q^2+q \equiv -1  \textup{mod} \ q^2+q+1$, the above equation can be simplified as
\begin{equation}\label{eq006}
{{r(2+ (\alpha+\beta)+q\beta)+qi-(q+1)j+k}\equiv 0\ (\textup{mod}\ q^2+q+1)}.
\end{equation}

For given $\alpha,\beta$, when ${0\leq i\leq \alpha, 0\leq j\leq \beta, 0\leq k\leq 2(q-1)-\alpha-\beta}  $, the minimal value of 
\[
{r(2+ (\alpha+\beta)+q\beta)+qi-(q+1)j+k}
\] is 
\[
m={r(2+ (\alpha+\beta)+q\beta)-(q+1)\beta},
\]
  and the maximal value is
\[
M={r(2+ (\alpha+\beta)+q\beta)+q\alpha+(2(q-1)-(\alpha+\beta))}.
\]
Thus
\begin{equation}\label{eq6}
M-m=(q-1)(2+\alpha)+q\beta.
\end{equation}
For the case $\alpha=\gamma=q-1, \beta=0$,
\begin{equation}\label{eq007}
M-m=q^2-1<q^2+q+1,
\end{equation}
that is there exists at most one pair of $i,k$ such that congruence equation (\ref{eq006}) can be  satisfied.

\begin{lemma}\label{bbl1}
Let $f(x)=x^r(x^{q-1}+a)\in \mathbb{F}_{q^3}[x]$, $1\leq r\leq q^2+q+1$. Then $f(x)$ is not a permutation binomial over $\mathbb{F}_{q^3}$ when $r\not= r_0q+1$ for some integer $1 \leq r_0\leq q$, here $a^{q^2+q+1}\not=-1$; and $f(x)$ is a permutation polynomial when $r=1$.
\end{lemma}

\begin{IEEEproof}
It can be verified that $f(x)=0$ has only one zero in $\mathbb{F}_{q^3}$ since $a^{q^2+q+1}\not=-1$.
Since the value range of $r$ is $1\leq r\leq q^2+q+1$, for integer $0\leq r_0\leq q$, let us consider the situation that
\[
2+r_0q\leq r\leq q+r_0q.
\]
Set $\beta =0$, then $\alpha=\gamma=q-1$ in the analysis above. Our aim is to prove that
\[
\sum\limits_{x\in \mathbb{F}_{q^3}}f(x)^{q-1+(q-1)q^2}\not= 0.
\]
Equation (\ref{eq006}) can be simplified as 
\begin{equation}\label{eq01}
r(2+\alpha)+qi+k=r(q+1)+qi+k \equiv 0\ (\textup{mod}\ q^2+q+1).
\end{equation}
And 
\[
2(q+1)+r_0(q^2+q)\leq r(q+1)\leq q^2+q+r_0(q^2+q).
\]
%thus
%\[
%2(q+1)-r_0 \leq r(q+1)-r_0(q^2+q+1)\leq q^2+q-r_0  
%\]
For $0\leq i,k\leq q-1$, we have $0\leq qi+k\leq q^2-1$. Since
\[
 q+2 \leq 2(q+1)-r_0  \leq r(q+1)-r_0(q^2+q+1) \leq q^2+q , 
\]
%\[
%\begin{array}{lll}
% q+2&=&2(q+1)-q \\
 % &\leq&2(q+1)-r_0 \\
  %                &\leq&r(q+1)-r_0(q^2+q+1)\\
   %               &\leq&q^2+q-r_0\\
    %              &\leq&q^2+q         
 %\end{array}
%\]
 there exist $i^*,k^*$ such that
\[
[r(q+1)-r_0(q^2+q+1)]+i^*q+k^*=q^2+q+1,
\] 
that is
\[
r(q+1)+i^*q+k^* \equiv 0\ (\textup{mod}\ q^2+q+1).
\]

According to (\ref{eq007}), %for given $\alpha, \beta$, 
there exists at  most one such $i^*,k^*$ for the above congruence equation to be satisfied. 
%So, for $N=(q-1)+(q-1)q$, when $f(x)^N$ is expanded, the 
For $N=q-1+(q-1)q^2$, by (\ref{eq002})
\[
\begin{array}{lll}
\sum\limits_{{x\in \mathbb{F}_{q^3}}}f(x)^N&=& \sum\limits_{{x\in \mathbb{F}_{q^3}}}f(x)^{\alpha+\beta q+\gamma q^2}\\

&=&
          
         -a^{(1-q)((2+\alpha)(1+q)+\beta q)}   \dbinom{\alpha}{i^*}\dbinom{\beta}{0}\dbinom{2(q-1)-\alpha-\beta}{k^*}a^{-i^*-q0-q^2k^*} \\

&=&
          
         -a^{q-q^2 -i^*-q^2k^*}   \dbinom{q-1}{i^*}\dbinom{q-1}{k^*}.  \\

 \end{array}
\]
From Lucas theorem, $\dbinom{q-1}{i^*}\dbinom{q-1}{k^*} \not\equiv 0 \ (\textup{mod}\ p)$. By Lemma 2, $f(x)$ is not a permutation polynomial for those values of $r$.

When $0\leq r_0\leq q$, the range of values of $r$ is % $r\in r_0q\leq r\leq q+r_0q$ are
\[
[2,q]\bigcup [2+q,q+q ]\bigcup [2+2q, q+2q]\bigcup \cdots \bigcup [2+q^2,q +q^2]. 
\] 
The values of $r$  that are missing are $1 $, and $q^2+q+1 $. Now let us consider these two cases. 

For $r=1$,  
\[
f(x)=x^q+ax,
\]
which is a linearized polynomial. Since $a^{q^2+q+1}\not= -1$, it is a permutation polynomial.

For $r=q^2+q+1$, for equation (\ref{eq01}) to be satisfied, set $i^*=k^*=0$. Equation (\ref{eq002}) becomes
\[
\begin{array}{lll}
\sum\limits_{{x\in \mathbb{F}_{q^3}}}f(x)^N=
          
         -a^{q-q^2 } \not= 0,  \\

 \end{array}
\]
and $f(x)$ is not a permutation binomial for this case.
\end{IEEEproof}

\begin{example}
Take $\mathbb{F}_q=\mathbb{F}_{2^3}, a=\omega^i$, where $\omega$ is a primitive element of the finite fields $
\mathbb{F}_{2^9}$, $2\leq r \leq 73, r\not=8r_0+1$ for some integer $r_0$, and  $1\leq i\leq 510$ is not a multiple of $7$. Using Magma, it can be checked that
\[
f(x)=x^{r+7}+ax^r
\]
is  not a  permutation polynomial  of $
\mathbb{F}_{2^9}$. Finite field  $\mathbb{F}_{3^2}$ with extension field $\mathbb{F}_{3^6}$, and   finite field $\mathbb{F}_7$ with extension field $\mathbb{F}_{7^3}$ are also verified for Theorem \ref{t01}.
\end{example}

In Theorem \ref{t01}, the values of $r$ that are not known for $f(x)$  to be a permutation polynomial are of the form $r_0q+1$. In the following proposition, many values of such $r$ are studied.

 \begin{proposition}\label{p01}
Let $f(x)=x^r(x^{q-1}+a)\in \mathbb{F}_{q^3}[x]$, where $\mathbb{F}_q$ is of odd characteristic. Then $f(x)$ is not a permutation binomial over $\mathbb{F}_{q^3}$ in the following two cases for $r= r_0q+1$ for some integer $1 \leq r_0\leq q$, here $a^{q^2+q+1}\not=-1$.

\begin{enumerate}
\renewcommand{\labelenumi}{$($\mbox{\roman{enumi}}$)$}
\item
If $r_0$ is an odd integer, then according to Lemma 1, $f(x)$ can not be a permutation polynomial as $\textup{gcd}(r=r_0q+1,q-1)\geq 2$;

%$1\leq r_0 \leq {{q-1}\over 2}$ is an odd integer, and the digits of $p$-expansions of ${{r_0-1}\over 2}$ and %$r_0$ are both not bigger than ${{p-1}\over 2}$;
\item
If $2\leq r_0 \leq {{q-1} }$ is an even integer, and the digits of $p$-expansions of ${{r_0-2}\over 2}$ is not bigger than ${{p-1}\over 2}$.
 
\end{enumerate}
\end{proposition}
 
\begin{IEEEproof}
 We only consider case (ii). Let $\alpha=\beta ={{q-1}\over 2},\gamma=q-1$. By equation (\ref{eq6}) 
\[
M-m=(q-1)(2+\alpha)+q\beta =q^2+{{q-3}\over 2}< q^2+q+1.
\]
So, there is at most one solution $i^*,j^*,k^*$ for the congruence relation (\ref{eq006}). And equation (\ref{eq006}) can be simplifed as 
\[
{r_0\over 2}q+{{q^2+q+2}\over 2}+qi-(q+1)j+k \equiv 0\ (\textup{mod}\ q^2+q+1).
\]
Now, let $i^*={{q-1}\over 2}, j^*={{r_0-2}\over 2}$,  then
\[
-{r_0\over 2}+1+k \equiv 0\ (\textup{mod}\ q^2+q+1),
\]
that is $k^*={r_0\over 2}-1$. And equation (\ref{eq002}) becomes
\[
\begin{array}{lll}
&&\sum\limits_{{x\in \mathbb{F}_{q^3}}}f(x)^{\alpha+\beta q+\gamma q^2}\\

 &=&
         -a^{{{q^2+ q+2}\over 2}-{{r_0\over 2}}q(q+1)}    \dbinom{{{q-1}\over 2}}{{{q-1}\over 2}}
\dbinom{{{q-1}\over 2}}{{ {r_0-2}\over 2}}\dbinom{q-1}{{r_0\over 2}-1} \\
&\not=& 0,   

 \end{array}
\]
by Lemma 3. That is $f(x)$ is not a permutation polynomial in this case.
\end{IEEEproof}

\begin{example}
Take $\mathbb{F}_q=\mathbb{F}_{3^2}, a=\omega^i$ where $\omega $ is a primitive element of $\mathbb{F}_{3^6}, 1\leq i \leq 727, i \neq 4 \ \text{mod} \ 8$. Let $r=9r_0+1$, $r_0=1,3,5,7,9$ or $r=19,37$, using Magma it can be verified that
\[
f(x)=x^{r+8}+ax^r
\]
is  not a  permutation polynomial over finite field $\mathbb{F}_{3^6}$. Finite field $\mathbb{F}_7$ with extension field $\mathbb{F}_{7^3}$ is also verified for Proposition \ref{p01}.
\end{example}

 \begin{remark}\label{r01}
According to Theorem \ref{t01}, Proposition \ref{p01}, and the fact that $f(x)$ is a permutation polynomial only possible in the case $\textup{gcd}(r_0+1,q-1)=1$, we guess that $f(x)$ is a permutation polynomial only when $r=1$.
\end{remark}
 
\begin{example}
Take $\mathbb{F}_q=\mathbb{F}_{13}, a=\omega^i$ where $\omega $ is a primitive element of $\mathbb{F}_{13^3}, 1\leq i \leq 2195, i \neq 6 \ \text{mod} \ 12$. Let $r=13r_0+1$, $r_0=4,6,10,12$, using Magma it can be verified that
\[
f(x)=x^{r+12}+ax^r
\]
is  not a  permutation polynomial over finite field $\mathbb{F}_{13^3}$. Finite field $\mathbb{F}_7$ with extension field $\mathbb{F}_{7^3}$  and finite field $\mathbb{F}_{3^2}$ with extension field $\mathbb{F}_{3^6}$ are also verified for the guess in Remark \ref{r01}.

\end{example}

\begin{remark}
If $\mathbb{F}_q$ is of even  characteristic, take $r_0=q, r=r_0q+1=q^2+1$, it can be found that
\[
f(x)=x^r(x^{q-1}+a)
\]
is a permutation of $\mathbb{F}_{q^3}$. So, we only consider the odd characteristic case for the guess in Remark \ref{r01}, and a complete determination is expected for both cases.
\end{remark}

Now, let us consider the case  $r=r_0q+1$ more for odd characteristic.
\begin{lemma}\label{bbl2}
Let $f(x)=x^r(x^{q-1}+a)\in \mathbb{F}_{q^3}[x]$, $1\leq r\leq q^2+q+1$. Then $f(x)$ is not a permutation binomial over $\mathbb{F}_{q^3}$ when $r=r_0q+1$  unless $p|r_0$, here $a^{q^2+q+1}\not=-1$, $q$ is a power of an odd prime and $1 \leq r_0\leq q$ is a positive integer.
%when $r\not= r_0q+1$ for some integer $1 \leq r_0\leq q$, here $a^{q^2+q+1}\not=-1$; and $f(x)$ is a permutation polynomial when $r=1$.
\end{lemma}
\begin{IEEEproof}
For Lemma \ref{lemma2}, we have
\begin{equation*}\label{bb1}
f(x)=x^{(r_0+1)q}+ax^{r_0q+1}
\end{equation*}
where $2\leq r_0\leq q-1$, since $r_0$ is even by Remark \ref{r01}.
Then for $1\leq N\leq q^3-2$
\begin{equation*}\label{bb2}
\begin{array}{lll}
f(x)^N &=&(x^{(r_0+1)q}+ax^{r_0q+1})^N\\
 &=&  \sum\limits_{n_1=0}^{N} \dbinom{N}{n_1} x^{(r_0+1)qn_1+(r_0q+1)n_2}a^{n_2} \\
 &=&  \sum\limits_{n_1=0}^{N} \dbinom{N}{n_1} x^{r_0qn_2+n_2+r_0qn_1+qn_1}a^{n_2} \\
 &=&  \sum\limits_{n_1=0}^{N} \dbinom{N}{n_1} x^{r_0qN+N+(q-1)n_1}a^{n_2} \\
 &=&  \sum\limits_{n_1=0}^{N} \dbinom{N}{n_1} x^{(r_0q+1)N+ (q-1)n_1}a^{n_2}, \\
 \end{array}
\end{equation*}
whose term has degree less than $q^3-1$   modular $x^{q^3}-x$ unless
\begin{equation*}\label{bb3}
{(r_0q+1)N+ (q-1)n_1} \equiv 0 \ \ (\textup{mod} \ \ q^3-1) 
\end{equation*}
where $n_2=N-n_1$.
Since $q-1$ is a divisor of $q^3-1$ and $\textup{gcd}(r_0+1,q-1)=1$ by Remark \ref{r01} , we have
\begin{equation*}\label{bb4}
 N=n_1+n_2=k(q-1)
\end{equation*}
for some integer $k$. Thus
\begin{equation}\label{bb5}
r_0qk+k+n_1 \equiv 0 \ \ (\textup{mod} \ \ q^2+q+1).
\end{equation}

First, let us assume that $r_0\leq {{q-1}\over 2}$. Take $k=q^2+q-1$, then
\begin{equation}\label{bb6}
N=k(q-1)=1+(q-2)q+(q-1)q^2.
\end{equation}
From equation (\ref{bb5}), we have
\begin{equation*}\label{bb7}
r_0(q^3+q^2-q)+q^2+q-1+n_1 \equiv 0 \ \ (\textup{mod} \ \ q^2+q+1),
\end{equation*}
which is equivalent to
\begin{equation*}\label{bb8}
r_0(1-q-1-q)-1-1+n_1 \equiv 0 \ \ (\textup{mod} \ \ q^2+q+1).
\end{equation*}
That is 
\begin{equation*}\label{bb9}
n_1 \equiv 2+2r_0q \ \ (\textup{mod} \ \ q^2+q+1),
\end{equation*}
and 
\begin{equation*}\label{bb10}
n_1 = 2+2r_0q +k_0(q^2+q+1) 
\end{equation*}
for some integer $k_0$. Comparing with equation (\ref{bb6}), we have $k_0=q-2$. Then
\begin{equation*}\label{bb11}
\begin{array}{lll}
n_1 &=&(2r_0+1+q-2)q +(q-2)q^2 \\
 &=&   (2r_0-1)q +(q-1)q^2.
 \end{array}
\end{equation*}
We have $2r_0-1\leq q-2$, and by Lemma \ref{lemma3}
\begin{equation*}\label{bb12}
\dbinom{q-2}{2r_0-1}
\end{equation*}
is zero when 
\begin{equation*}\label{bb13}
 2r_0-1  \equiv p-1 \ \ (\textup{mod} \ \ p) 
\end{equation*}
that is $r_0$ is a multiple of $p$, since $p$ is odd.

Second, let us assume that $r_0> {{q-1}\over 2}$. Take $k=q^2-q+1$, then
\begin{equation}\label{bb14}
N=k(q-1)=(q-1)+ q+(q-2)q^2.
\end{equation}
From equation (\ref{bb5}), we have
\begin{equation*}\label{bb15}
r_0(q^3-q^2+q)+q^2-q+1+n_1 \equiv 0 \ \ (\textup{mod} \ \ q^2+q+1),
\end{equation*}
which is equivalent to
\begin{equation*}\label{bb16}
r_0(1+q+1+q)-q-q+n_1 \equiv 0 \ \ (\textup{mod} \ \ q^2+q+1).
\end{equation*}
That is 
\begin{equation*}\label{bb17}
n_1 \equiv -2r_0-(2r_0-2)q \ \ (\textup{mod} \ \ q^2+q+1),
\end{equation*}
and 
\begin{equation*}\label{bb118}
\begin{array}{lll}
n_1 &= & -2r_0-(2r_0-2)q  +k_0(q^2+q+1) \\
 &=&   -(2r_0-q)+(1-(2r_0-q))q-q^2  +k_0(q^2+q+1) 
 \end{array}
\end{equation*}
for some integer $k_0$. Comparing with equation (\ref{bb14}), we have $k_0=2r_0-q$. Then
\begin{equation*}\label{bb19}
\begin{array}{lll}
n_1 =q+(2r_0-q-1)q^2.\\
 %&=&   (2r_0-1)q +(q-1)q^2.
 \end{array}
\end{equation*}
We have 
\[
 -2=q-1-q-1<2r_0-1 -q \leq 2(q-1)-1-q=q-3.
\]  
By Lemma \ref{lemma3}
\begin{equation*}\label{bb20}
\dbinom{q-2}{2r_0-1-q}
\end{equation*}
is zero when 
\begin{equation*}\label{bb21}
 2r_0-1-q  \equiv p-1 \ \ (\textup{mod} \ \ p), 
\end{equation*}
that is $r_0$ is a multiple of $p$, since $p$ is odd.  By Lemma \ref{lemma2}, we get the results.  
\end{IEEEproof}

The following lemma is about the case when $p|r_0$.
\begin{lemma}\label{bbl3}
Let $f(x)=x^r(x^{q-1}+a)\in \mathbb{F}_{q^3}[x]$, $1\leq r\leq q^2+q+1$. Then $f(x)$ is not a permutation binomial over $\mathbb{F}_{q^3}$ when $r=r_0q+1$  and $p|r_0$, here $a^{q^2+q+1}\not=-1$,  $q$ is a power of an  odd prime  and $1 \leq r_0\leq q$ is a positive integer.
%when $r\not= r_0q+1$ for some integer $1 \leq r_0\leq q$, here $a^{q^2+q+1}\not=-1$; and $f(x)$ is a permutation polynomial when $r=1$.
\end{lemma}

\begin{IEEEproof}
%First let's assume that $r_0> {{q-1}\over 2}$.
As in the analysis of Lemma \ref{bbl2}, let us take $k=q^2- 1$, then
\begin{equation}\label{bb22}
N=k(q-1)=1+ (q-1)q+(q-2)q^2.
\end{equation}
From equation (\ref{bb5}), we have
\begin{equation*}\label{bb23}
r_0(q^3-q )+q^2- 1+n_1 \equiv 0 \ \ (\textup{mod} \ \ q^2+q+1),
\end{equation*}
which is equivalent to
\begin{equation*}\label{bb24}
r_0(1-q)-q-1-1+n_1 \equiv 0 \ \ (\textup{mod} \ \ q^2+q+1).
\end{equation*}
That is 
\begin{equation*}\label{bb25}
n_1 \equiv 2-r_0+(r_0+1)q \ \ (\textup{mod} \ \ q^2+q+1),
\end{equation*}
and 
\begin{equation*}\label{bb26}
\begin{array}{lll}
n_1=2-r_0+(r_0+1)q \ +k_0(q^2+q+1).\\
% &=&   -(2r_0-q)+(1-(2r_0-q))q-q^2  +k_0(q^2+q+1).
 \end{array}
\end{equation*}
for some integer $k_0$.

Fist we consider  the case that $r_0<{{q+1}\over 2}$. 
Comparing with equation (\ref{bb22}), we have %$k_0= r_0-2$ or . Then
\begin{equation}\label{bb27}
\left\{
\begin{array}{lll}
n_1 &=& (2r_0-1)q+(r_0-2)q^2 \ \ \text{when} \ \ k_0=r_0-2,\\
n_1 &=& 1+2r_0 q+(r_0-1)q^2  \ \ \  \  \text{when} \ \ k_0=r_0-1.
 \end{array}
\right.
\end{equation}
For the second equality of the above equation 
\[
r_0-1 \equiv p-1 \ \ (\textup{mod} \ \ p) 
\]
since $p|r_0$.
That is,
\begin{equation*}\label{bb28}
 \dbinom{q-2}{r_0-1 } \equiv 0 \ \ (\textup{mod} \ \ p).
\end{equation*}
by Lemma \ref{lemma3}. 
For the first equality of equation (\ref{bb27}) 
\begin{equation}\label{bb29}
\dbinom{N}{n_1}  \not\equiv 0 \ \ (\textup{mod} \ \ p). 
\end{equation}

Second we consider the case that $r_0\geq{{q+1}\over 2}$. 
 We have %$k_0= r_0-2$ or . Then
\begin{equation}\label{bb30}
\left\{
\begin{array}{llll}
n_1 &=& (2r_0-1-q)q+(r_0-1)q^2 \ \ &\text{when} \ \ k_0=r_0-2\\
n_1 &=& 1+(2r_0 -q)q+r_0 q^2  \ \ \  \ \ \  \ &\text{when} \ \ k_0=r_0-1.
 \end{array}
\right.
\end{equation}
For the first equality of the above equation 
\[
r_0-1 \equiv p-1 \ \ (\textup{mod} \ \ p). 
\]
That is,
\begin{equation*}\label{bb31}
 \dbinom{N}{n_1} \equiv \dbinom{q-2}{r_0-1 } \equiv 0 \ \ (\textup{mod} \ \ p). 
\end{equation*}
From the second equality of equation (\ref{bb30}), we also get equation (\ref{bb29}). By Lemma \ref{lemma2}, we get our results.  
%\begin{equation}\label{bb32}
%\dbinom{N}{n_1}  \not\equiv 0 \ \ (\textup{mod} \ \ p).   
%\end{equation}
\end{IEEEproof}

Combining   Lemma \ref{bbl1}, Lemma \ref{bbl2} and Lemma \ref{bbl3}, we have
 \begin{theorem}\label{bbt1}
Let $f(x)=x^r(x^{q-1}+a)\in \mathbb{F}_{q^3}[x]$, $1\leq r\leq q^2+q+1$. Then $f(x)$ is a permutation binomial over $\mathbb{F}_{q^3}$ if and only if $r= 1$, here $a^{q^2+q+1}\not=-1$ and $q$ is a power of an  odd prime.
%when $r\not= r_0q+1$ for some integer $1 \leq r_0\leq q$, here $a^{q^2+q+1}\not=-1$; and $f(x)$ is a permutation polynomial when $r=1$.
\end{theorem}

\section{The case of finite field  $\mathbb{F}_{q^e}$}\label{Section III}

For $a\in \mathbb{F}_{q^e}^*$, in this section, we consider the permutation properties of the binomial $f(x)=x^r(x^{q-1}+a)$. Different from last section, only large values of $e$ such that $f(x)$ is not a permutation polynomial are quantified.

 \begin{theorem}\label{MR001}
Let $1 <r<q^{e\over 4}-q+3$ be an integer, and $\mathbb{F}_{q^e}$ is a finite field %of odd characeristic
 with $q^e$ elements, and $a\in \mathbb{F}_{q^e}^*, q\geq 6$, then $f(x)=x^r(x^{q-1}+a)$ is not a permutation polynomial over $\mathbb{F}_{q^e}$. 
\end{theorem}
 
\begin{IEEEproof}
 In fact, by Lemma 1 $\textup{gcd}(r,q-1)\not= 1$ means that $f(x)$ is not a permutation polynomial, so let us consider the case that $\textup{gcd}(r,q-1)=1$.
Set \[
F(X,Y)={{f(X)-f(Y)}\over {X-Y}}={{X^{r+q-1}-Y^{r+q-1}}\over {X-Y}}+a{{X^{r}-Y^{r}}\over {X-Y}}.
\]
We claim that $F(X,Y)$ is absolutely irreducible over $\bar{\mathbb{F}}_q$, by considering two cases. Let
\[
\bar{F}(X,Y,Z)= {{X^{r+q-1}-Y^{r+q-1}}\over {X-Y}}+a{{X^{r}-Y^{r}}\over {X-Y}}Z^{q-1} 
\]
be the homogenization of $F(X,Y)$. It is only necessary  to show that  
\[
\bar{F}(X,1,Z)= {{X^{r+q-1}-1}\over {X-1}}+a{{X^{r}-1}\over {X-1}}Z^{q-1}
\] is irreducible in $\bar{\mathbb{F}}_q[X,Z]$. 

First,  consider the case that $r \equiv 1 \ \textup{mod} \ p$. Then 
the derivative of $x^r-1$ is $rx^{r-1}$ which is not zero, and $\textup{gcd}(x^r-1,rx^{r-1})=1$. So, $x^r-1$ has no  multiple roots.  Assume the contrary that 
\[
\bar{F}(X,1,Z)=f(X,Z)g(X,Z),
\]
where 
\[
f(X,Z)=\sum\limits_{i=0}^{m}f_i(X)Z^i
\]
\[
g(X,Z)=\sum\limits_{i=0}^{n}g_i(X)Z^i.
\]
Since 
\[
\textup{gcd}(X^{q+r-1}-1,X^r-1)=X^{\textup{gcd}(q+r-1,r)}-1=X-1,
\]
   $f(X,Z)$ and $g(X,Z)$ must both have positive power in $Z$, that is $ m,n\geq 1$. Since $ r>1$, 
\[
{{X^{r}-1}\over {X-1}}
\]
 has $ r-1>0$ different roots in $\bar{\mathbb{F}}_q$. Let $\xi$ be one such root, note that 
\[
{{X^{r}-1}\over {X-1}}=f_m(X)g_n(X),
\]
$\xi$ can only be a root of $f_m(X)$ or $g_n(X)$. Let us assume that $f_m(\xi)=0$, then $g_n(\xi)\not=0$
. Thus
\[
\bar{F}(\xi,1,Z)=f(\xi,Z)g(\xi,Z)={{\xi^{r+q-1}-1}\over {\xi-1}}=f_0(\xi)g_0({\xi})
\]
is a nonzero constant, since 
\[
\textup{gcd}( {{X^{r+q-1}-1}\over {X-1}},{{X^{r}-1}\over {X-1}})=1,
\]
and $\xi$ is not a root of ${{X^{r+q-1}-1}\over {X-1}}$.
But
\[
g(\xi,Z)=\sum\limits_{i=0}^{n-1}g_i(\xi)Z^i+g_n(\xi)Z^n
\]
has degree $ n\geq 1$ in $Z$, and $f_0(\xi)\not= 0$, contradiction.

Second, let us consider the case that $r \not\equiv 1 \ \textup{mod} \ p$. Then 
the derivative of $x^{r+q-1}-1$ is $(r-1)x^{r+q-2}$ which is not zero, and $\textup{gcd}((r-1)x^{r+q-2},x^{r+q-1}-1)=1$. So, $x^{r+q-1}-1$ has no  multiple roots. 
Arguing as in the first case
\[
{{X^{r+q-1}-1}\over {X-1}}
\]
 has $ r+q-2>1$ different roots in $\bar{\mathbb{F}}_q$. Let $\eta$ be one of the roots, 
then 
\[
{{X^{r+q-1}-1}\over {X-1}}=f_0(X)g_0(X),
\]
$\eta$ can only be a root of $f_0(X)$ or $g_0(X)$, not both. Let us assume that $f_0(\eta)=0$, then $g_0(\eta)\not= 0$ and $f_m(\eta)\not=0$. Then
\[
\bar{F}(\eta,1,Z)= a{{\eta^{r}-1}\over {\eta-1}}Z^{q-1}
\]
is a nonzero monomial in $Z$. But
\[
\bar{F}(\eta,1,Z)=f(\eta,Z)g(\eta,Z), 
\]
with $g(\eta,Z)$ satisfying $g_0(\eta),g_m(\eta)\not=0$, which is impossible.

So, $F(X,Y)$ is absolutely irreducible, and $X-Y \nmid F(X,Y)$.

 Now,
\[
F(X,X)=(r+q-1)X^{r+q-2}+ar X^{r-1}=X^{r-1}((r-1)X^{q-1}+ar)
\]
  has at most $q$ roots over $\mathbb{F}_{q^e}$.
  $F(X,Y)$ has degree $d=r+q-2$. According to the Hasse-Weil bound \cite{L001,S001,W0001}, the number of zeros of $F(X,Y)$ has   lower bound 
\[
|V_{\mathbb{F}_{q^e}}(F(X,Y))|\geq q^e-(d-1)(d-2)q^{e/2}-{1\over 2}d(d-1)^2
-d-2.
\]
Let $\lambda$ denote the larger root of 
\[
x^2-(d-1)(d-2)x-{1\over 2}d(d-1)^2-d-q-2.
\]
Then 
\[
\begin{array}{lll}
 \lambda&=&{1\over 2}[(d-1)(d-2)+\sqrt{(d-1)^2(d-2)^2+2d(d-1)^2+4d+4q+8}]\\
                  &=&{1\over 2}[d^2-3d+2+\sqrt{d^4-4d^3+7d^2-6d+4q+12}]\\
                  &\leq&{1\over 2}[d^2-3d+2+\sqrt{(d^2-d)^2}]\\
                  &=&d^2-2d+1=(d-1)^2\\
                  &<&q^{e\over 2}.
 \end{array}
\]
Therefore
\[
|V_{\mathbb{F}_{q^e}}(F(X,Y))|-q\geq q^e-(d-1)(d-2)q^{e/2}-{1\over 2}d(d-1)^2-q
-d-2>0.
\]
That is $F(X,Y)$ has more than $q$ zeros in $\mathbb{F}_{q^e}^2$, and only at most $q$ of them are on the line $X=Y$, thus $F(X,Y)$ has a zero with $X\not=Y$. Then $f(x)$ is not a permutation polynomial of  $\mathbb{F}_{q^e}$.
\end{IEEEproof}

\section{Conclusion}\label{SecIV}

In this paper, it is shown that $f(x)=x^r(x^{q-1}+a)$ is a permuation polynomial over $\mathbb{F}_{q^3}$ when $r=1$, and it is not a PP almost all the other cases, and we guess that $f(x)$ can be a permutation binomial  only when $r=1$ for odd characteristic. For $e$ large enough than $r$, we find that $f(x)$ is not a permutation polynomial over $\mathbb{F}_{q^e}$. 
 
\section*{Data Availability}\label{secVI}

The data used to support the findings of this study are available from the corresponding author upon request.

\section*{Conflict of Interests} \label{secVII}

%Conflict of Interests

The author declares that there is no conflict of interests regarding the publication of this paper.

 \section*{Acknowledgment}

 The author would like to thank the anonymous referees for helpful suggestions and comments.

%.............

\end{document}